\documentclass{article} 
\usepackage{iclr2027_conference,times}

\usepackage{amsmath,amsfonts,bm}

\def\eqref#1{equation~\ref{#1}}

\def\1{\bm{1}}

\DeclareMathAlphabet{\mathsfit}{\encodingdefault}{\sfdefault}{m}{sl}
\SetMathAlphabet{\mathsfit}{bold}{\encodingdefault}{\sfdefault}{bx}{n}

\usepackage{microtype}
\usepackage{graphicx}
\usepackage{subcaption}
\usepackage{hyperref}
\usepackage{xurl}
\hypersetup{hidelinks}
\usepackage{fontawesome5}
\usepackage{url,graphicx,booktabs,multirow,array}
\usepackage{fvextra}
\usepackage[table]{xcolor}
\usepackage{booktabs}
\usepackage{tcolorbox}
\usepackage{multirow}
\usepackage[table]{xcolor}
\usepackage{graphicx}
\usepackage{pifont}
\usepackage{threeparttable}

\title{CharDuplex: Building Character-Consistent Full-Duplex Spoken Dialogue Models}

\author{Donghang Wu$^{1,2,*}$ \ Yisi Liu$^{3,}$\thanks{Equal contribution} \quad  Chen Chen$^{1,\dag}$ \ \textbf{Hexin Liu$^{1}$\quad Eng Siong Chng$^{1,}$\thanks{\small{Corresponding authors: \texttt{chen1436@e.ntu.edu.sg; ASESChng@ntu.edu.sg} } } } \\
$^1$Nanyang Technological University \quad $^2$AI Singapore \quad $^3$University of California, Berkeley  
}

\iclrfinalcopy 
\begin{document}

\maketitle

\begin{abstract}
Full-duplex speech models are moving voice interaction beyond conventional turn-taking, yet natural conversation is shaped not only by when an agent speaks, but also by how it behaves as a conversational character. We present CharDuplex, a character-driven full-duplex speech model that combines real-time spoken interaction with persona-conditioned behavior. We first adapt GLM-4-Voice to an always-on dual-stream architecture and train the model for full-duplex conversation. Then a fully automated pipeline constructs character-conditioned dialogue data from open-source character descriptions for character-conditioned supervised fine-tuning. The model is further refined with the proposed FDGym, where an LLM-simulated user dynamically interacts with the model, enabling reinforcement learning over evolving multi-turn interactions. On SpeechRole-Eval, CharDuplex achieves the highest average score among the evaluated open-source models, while remaining competitive with closed-source systems. It also demonstrates competitive general speech intelligence and strong full-duplex interaction capabilities. CharDuplex demonstrates a practical training recipe for building full-duplex voice assistants that are not only interactive, but also character-consistent.
\end{abstract}

\section{Introduction}
Recent advances in Full-Duplex Spoken Dialogue Language Models (SDLMs) have brought spoken dialogue closer to continuous, real-time conversation. By processing incoming speech alongside response generation, full-duplex models allow listening and speaking to occur concurrently rather than exclusively in alternating turns \citep{moshi, chronological, duplexsla, salmonnomni, flair}. However, a natural conversation also depends on responses that reflect the specific character. A speech model should convey that character’s personality and perspective throughout the conversation, rather than respond uniformly as a general-purpose assistant. 

Character-consistent dialogue requires responses to address the user's input
while remaining compatible with the prescribed character's knowledge,
personality, and background. Knowledge constrains what the model can plausibly
claim, while personality and perspective shape how it interprets and answers
questions \citep{speechrole, omnicharacter}. In multi-turn interaction, these
constraints must persist across follow-up questions, corrections, and topic
shifts without contradicting earlier replies. Existing full-duplex work has
begun to incorporate such role conditioning. PersonaPlex, for example, uses
text-based role prompts together with speech-based voice prompts
\citep{personaplex}. Our focus is on strengthening character consistency in the response content of SDLMs across evolving multi-turn full-duplex interactions, where responses must remain
grounded in the prescribed character and coherent with the dialogue history.

Building such a full-duplex model requires training data that connect character descriptions to multi-turn spoken conversations. The supervision must capture how a character responds to different questions and dialogue histories, with speech inputs and target responses suitable for full-duplex training. Supervised training data, however, represent conversations whose continuations have already been determined. During live conversation, a user may question a claim, request an explanation, or ask a new follow-up question based on what the model has actually said. The model must address these reactions while remaining consistent with the same character. In full-duplex interaction, the user can also interrupt a response while the model is speaking, rather than waiting for it to finish. Therefore, character-consistent post-training in realistic multi-turn interaction requires response-dependent rollouts, where subsequent user behavior is conditioned on what the model actually generates. 

In this paper, we present \textbf{CharDuplex}, a character-driven full-duplex speech model built on GLM-4-Voice \citep{glm4voice}. We first establish its full-duplex capabilities through architectural adaptation and training on duplex conversational data, then construct synthetic multi-turn conversations from open-source character descriptions and convert them into speech data for character-conditioned Supervised Fine-Tuning (SFT). Finally, we develop \textbf{Full-Duplex Gym (FDGym)}, a real-time interactive training framework that connects the SDLM to a user simulator powered by a large language model (LLM). The simulator reacts to the model's incrementally available outputs and generates user speech online, forming a closed-loop, streaming conversation. Reinforcement Learning (RL) over these dynamically generated conversations further refines character-related response quality. CharDuplex substantially outperforms full-duplex SDLMs, especially PersonaPlex, across all four evaluated dimensions of SpeechRole-Eval \citep{speechrole} while maintaining strong performance on Full-Duplex-Bench \citep{fullduplex-bench} and VoiceBench \citep{voicebench}, demonstrating broad competence in both full-duplex interaction and general speech intelligence.

Our contributions are threefold:

\begin{itemize}
    \item We introduce CharDuplex, a full-duplex SDLM that combines character-consistent response generation with strong general speech intelligence and full-duplex interaction capabilities.
    \item We develop a fully automated pipeline for constructing character-conditioned full-duplex dialogue data, with dialogue generation and verification, speech-quality control, and synchronized interaction-event construction.
    \item We introduce FDGym, a response-dependent interactive RL framework in which subsequent simulated user speech is generated online from the model’s progressively available responses, enabling closed-loop multi-turn optimization of character consistency while preserving general conversational capabilities.
\end{itemize}

\section{Related Work}

\subsection{Full-Duplex Spoken Dialogue Models}

Full-duplex SDLMs enable concurrent listening and speaking rather than
alternating between fixed conversational turns \citep{voicemem, flair, soulx}. Moshi models user and system
speech in parallel streams \citep{moshi}, while subsequent systems explore
different architectural designs, including the frozen-LLM formulation of
Freeze-Omni \citep{freezeomni} and the codec-free SALMONN-omni
\citep{salmonnomni}. Recent work further improves the intelligence of
full-duplex models through temporally structured or latent reasoning
\citep{chronological,flair}. These studies primarily target general interaction
and reasoning capabilities; CharDuplex instead focuses on maintaining character
consistency while preserving these general capabilities.

\subsection{Character-Conditioned Spoken Interaction}

Character modeling has recently been extended from text dialogue to spoken
interaction. OmniCharacter models linguistic personality and vocal traits
\citep{omnicharacter}, while SpeechRole provides large-scale training data and
evaluation for spoken role-playing \citep{speechrole}. Most closely related to our work,
PersonaPlex conditions full-duplex conversation on textual role descriptions
and voice prompts \citep{personaplex}, while SteerDuplex considers broader
steerability over persona, tone, and speaking style \citep{steerduplex}.
CharDuplex focuses on character-consistent response generation across multi-turn
full-duplex interaction, supported by automated data construction and
interactive post-training.

\subsection{Reinforcement Learning for Full-Duplex SDLM}
Reinforcement learning has increasingly been used to address interaction-level
behaviors that are difficult to optimize through token-level supervised learning.
SALMONN-omni applies preference optimization after supervised training to improve
context-dependent barge-in and backchannel handling \citep{salmonnomni}.
ORISE further introduces online policy optimization with automatically derived
speech-based rewards \citep{orise}.
DuplexPO decouples
\textit{when to speak} from \textit{what to say}, optimizing selected interaction
windows with factorized rewards for turn initiation, backchanneling, yielding, and
participation \citep{decoupling}. 
These approaches substantially advance RL-based optimization of full-duplex
interaction, but their rollouts are primarily constructed from prerecorded speech, or predetermined conversational contexts.
However, different responses produced by the model should induce different subsequent user
utterances. This distinction becomes important for multi-turn character interaction. FDGym therefore introduces an LLM-based user simulator inside the rollout loop,
allowing character consistency to be optimized under conversational contexts that emerge
from the model's own behavior.
\begin{figure*}[t]
\centering
\includegraphics[width=12.5cm]{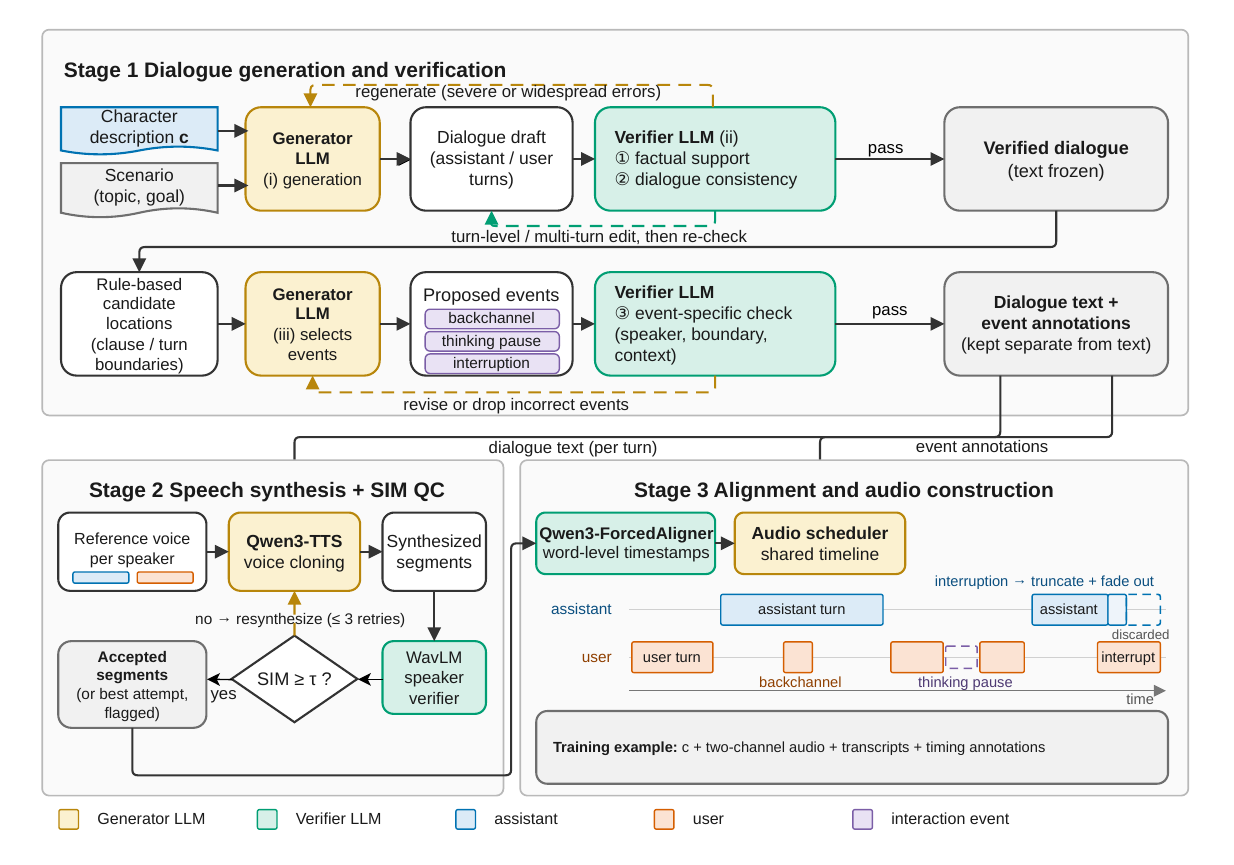}
\caption{
Overview of the automated character-conditioned data construction pipeline.
Stage 1 alternates a generator and verifier to generate, validate, repair, and
annotate multi-turn character dialogues. Stage 2 synthesizes speaker-conditioned
speech with speaker-similarity-based quality control. Stage 3 uses forced
alignment to realize pauses, backchannels, and interruptions on synchronized
user and assistant audio channels.
}
\label{fig:data_construction}
\vspace{-10pt}
\end{figure*}
\vspace{-5pt}
\section{Method}
\label{sec:method}

CharDuplex is a character-driven full-duplex SDLM built on GLM-4-Voice \citep{glm4voice}. Its construction comprises full-duplex adaptation and supervised training, character-conditioned SFT, and interactive RL with FDGym. We first describe the automated character-conditioned data construction pipeline, then introduce the model architecture, followed by the SFT and RL objectives.

\subsection{Character-Conditioned Data Construction}
\label{sec:character_data}

Character-conditioned speech data are constructed using character descriptions from the training splits of open-sourced datasets \citep{omnicharacter,speechrole}. The pipeline integrates an automated generator--verifier process for character-conditioned text dialogue generation, speech synthesis with speaker-similarity-based quality control, and forced alignment for constructing synchronized two-channel dialogue audio. The whole pipeline is shown in Figure \ref{fig:data_construction}.

\paragraph{Stage 1: Automated character-conditioned dialogue generation and verification.}
Stage 1 follows an automated generator--verifier loop that alternates generation, verification, and targeted correction. Given a character description and a sampled conversational scenario, the generator first produces a multi-turn dialogue conditioned on the prescribed character. The verifier then checks the dialogue with task-specific prompts for factual grounding with respect to the character description and dialogue history, as well as logical consistency across turns. Local defects trigger turn- or span-level correction followed by re-verification, whereas broader failures trigger dialogue regeneration. Once the dialogue content passes verification, the generator then identifies the locations for events including backchannels, user thinking pauses, and interruptions, and the verifier checks the resulting annotations before they are passed to the subsequent speech synthesis stages. Detailed generation, verification, repair, and data construction prompts are provided in Appendix~\ref{sec:app_char_data_generation}.

\paragraph{Stage 2: Speech synthesis.}
We use Qwen3-TTS to synthesize speech from each speaker's dialogue text \citep{qwen3-tts}, using a distinct reference recording for each speaker throughout the conversation. After that, a WavLM-based speaker verifier \citep{wavlm} from SEED-TTS-Eval \citep{seed-tts} measures similarity between synthesized and reference audio. Segments below a threshold of $0.5$ are resynthesized within a fixed retry budget; if none passes, the highest-scoring attempt is retained and marked.

\paragraph{Stage 3: Speech alignment and audio construction.}
We use Qwen3-ForcedAligner to align each synthesized segment with its text to obtain word-level timestamps \citep{qwen3-asr}. An audio scheduler uses these timestamps to place speech in separate user and assistant channels on a shared timeline. It inserts silence for user thinking pauses and overlays backchannels on ongoing speech. User interruptions start at annotated word boundaries; after a short overlap, the assistant audio is truncated and faded out, and the discarded continuation is removed from its transcript. Each training example retains the character description, two-channel audio, transcripts, and temporal annotations. For SFT, the description supplies the system prompt, the user audio supplies the speech input, and the assistant transcript and speech boundaries determine the target response stream.

\begin{figure*}[t]
\centering
\includegraphics[width=13cm]{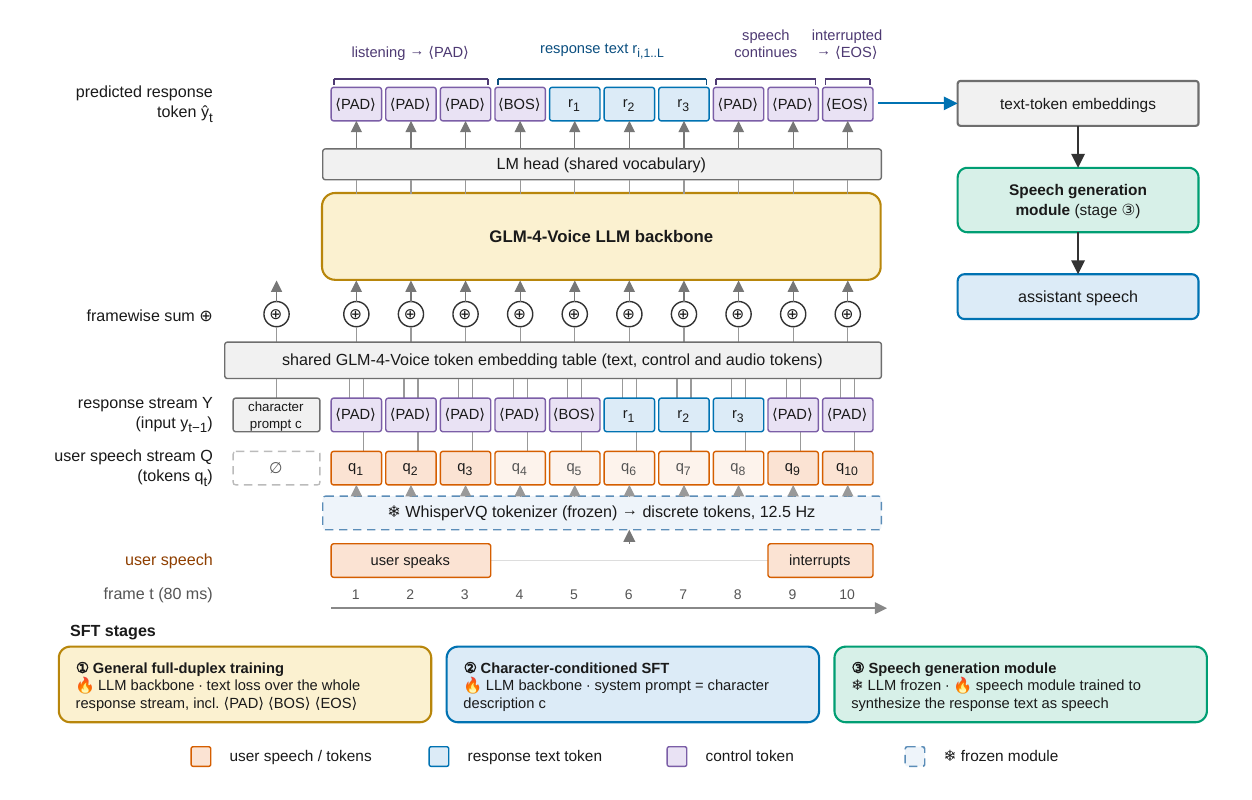}
\caption{
Overview of the CharDuplex architecture and supervised training.
GLM-4-Voice is reformulated into synchronous, always-on user-speech and
response-text streams that share the token embedding and causal backbone.
General full-duplex training is followed by
character-conditioned SFT and speech-generation training.
}
\label{fig:architecture}
\vspace{-5pt}
\end{figure*}
\subsection{Full-Duplex Model Architecture}
\label{sec:architecture}

\paragraph{Concurrent speech and response streams.}
We reformulate GLM-4-Voice's serialized conversation stream into two
synchronous, always-on streams: a user speech stream
$Q=(q_1,\ldots,q_T)$ represented by frozen WhisperVQ tokens at 12.5 Hz,
and a response stream $Y=(y_1,\ldots,y_T)$ containing text and control tokens
on the same frame timeline. Their embeddings are combined framewise and fed
to the causal backbone. The model architecture is shown in Figure \ref{fig:architecture}. Given the character description $c$ as the system prompt,
response generation is formulated as
\begin{equation}
p_\theta(Y\mid Q,c)
=
\prod_{t=1}^{T}
p_\theta(y_t\mid c,q_{<t},y_{<t}).
\end{equation}
The response stream is autoregressive, while the user stream is continuously
supplied by incoming audio.

\paragraph{Turn-level text--speech alignment.}
Response text and speech are aligned at the turn level rather than requiring
individual text tokens to coincide with spoken words. The response stream emits
\texttt{<PAD>} tokens while listening, followed by a \texttt{<BOS>} token and the response text.
After text generation finishes, the model continues to predict \texttt{<PAD>} tokens until the corresponding
speech ends, followed by an \texttt{<EOS>} token. For a response containing $L_i$ text tokens, the stream around turn $i$ takes the form
\begin{equation}
\begin{aligned}
    Y^{(i)}=\bigl[
    &\underbrace{
        \langle\mathrm{PAD}\rangle,\ldots,\langle\mathrm{PAD}\rangle
    }_{\text{before the response}},
    \langle\mathrm{BOS}\rangle,
    r_{i,1},\ldots,r_{i,L_i},\\
    &\underbrace{
        \langle\mathrm{PAD}\rangle,\ldots,\langle\mathrm{PAD}\rangle
    }_{\text{speech continues}},
    \langle\mathrm{EOS}\rangle
    \bigr],
\end{aligned}
\label{eq:turn_level_alignment}
\end{equation}
where $r_{i,j}$ denotes the $j$-th text token of response $i$. 

When the user interrupts an ongoing response, the model emits $\langle\mathrm{EOS}\rangle$ to terminate the current turn and stop its speech output, without waiting for the remaining text or speech to finish. The response stream then returns to PAD prediction, while the user speech stream remains active. This construction maintains continuous dual-stream processing while allowing a spoken turn to end either through normal completion or interruption.

\subsection{Supervised Fine-Tuning}

Supervised training comprises three stages: general full-duplex training,
character-conditioned SFT, and speech-generation-module training. In the first
two stages, the LLM backbone is optimized over the complete response stream
using next-token prediction \citep{ntp}:
\begin{equation}
\mathcal{L}_{\mathrm{text}}
=
-\sum_{t=1}^{T}
\log p_{\theta}
\left(
y_t \mid c, q_{\leq t}, y_{<t}
\right),
\end{equation}
where $c$ denotes the system-prompt context when present. The WhisperVQ tokenizer remains frozen.

The first stage trains the adapted GLM-4-Voice on general full-duplex
conversations to establish response generation under the dual-stream
representation and turn-level alignment. Data statistics,
the generation toolkit, and configurations are provided in Section 4 and Appendix B. Starting from this model, the second
stage continues training on the character-conditioned conversations constructed
in Section \ref{sec:character_data}, where the character description is provided as the system
prompt $c$. The same objective is retained, so the response is conditioned
jointly on the character description, incoming user speech, and preceding
response tokens.

Finally, the LLM backbone is frozen and the speech-generation module is trained
separately. An autoregressive Transformer predicts speech codec tokens
conditioned on the response text, and the predicted tokens are converted to
waveforms using the streaming flow-matching decoder of CosyVoice2 \citep{cosyvoice2}.


\begin{figure*}[t]
\centering
\includegraphics[width=14cm]{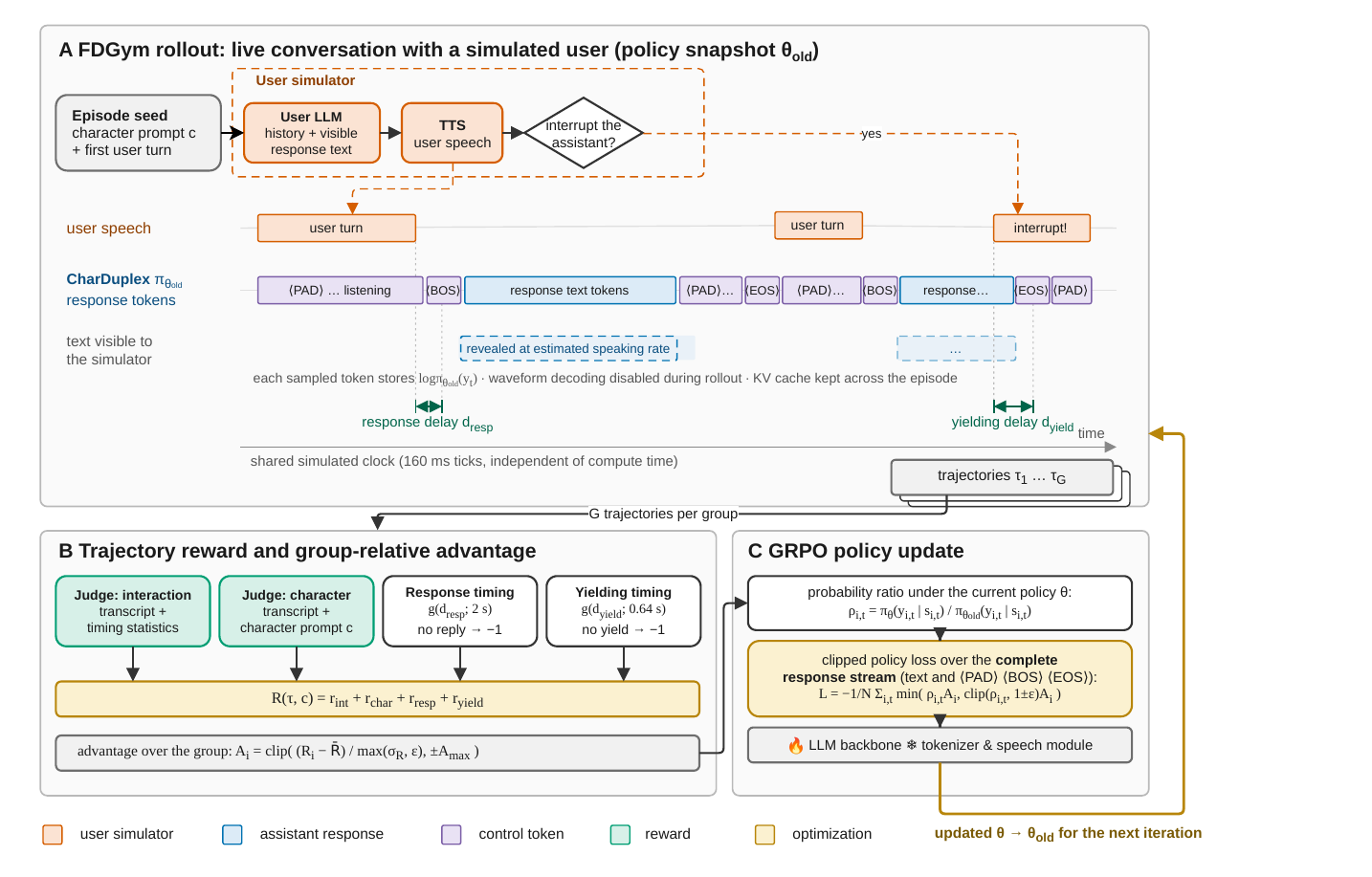}
\caption{
Overview of FDGym for real-time interactive RL.
(A) CharDuplex interacts with an LLM-based simulated user whose subsequent
speech depends on the model's progressively available responses.
(B) Each multi-turn trajectory receives character, interaction, and timing
rewards, from which a group-relative advantage is computed.
(C) GRPO updates the response policy using the sampled trajectory, and the
updated model is used for the next rollout iteration.
}
\label{fig:fdgym}
\vspace{-5pt}
\end{figure*}

\subsection{Real-Time Interactive Reinforcement Learning}
\label{sec:interactive_rl}

RL rollouts with prerecorded user continuations cannot capture how a model's replies affect subsequent user input \citep{decoupling, orise}. We develop FDGym, a streaming training environment that couples CharDuplex with an LLM-based user simulator and a dialogue evaluator. The pipeline is shown in Figure \ref{fig:fdgym}. The simulator generates user speech in response to the model's unfolding replies, while the evaluator provides character and interaction feedback. Starting from the SFT model, we alternate dynamic dialogue sampling and policy updates.

\paragraph{Full-duplex rollout generation.}
Each rollout starts with a character prompt and an initial user turn. FDGym
represents dialogue time using fixed-duration simulation ticks, with each tick
corresponding to 160\,ms \citep{tau-voice}. A tick serves as the basic
synchronization step of the full-duplex interaction: the user side supplies the speech available within the
current interval, while CharDuplex processes the accumulated incoming audio and
advances its response stream. The resulting partial interaction state is then
carried into the next tick, allowing user speech and model responses to evolve
incrementally on the same simulated timeline.

The user simulator generates subsequent utterances from the dialogue history
and the model's progressively available response, with Qwen3-TTS synthesizing
the corresponding user speech. During an ongoing model response, only the
portion that would have become audible by the current tick is exposed to the
simulator, according to a speaking rate estimated from the SFT data. The
simulator can therefore react to the response as it unfolds, including deciding
to interrupt it. During rollout, model-side waveform decoding is disabled and
the KV cache is retained throughout the rollout. The resulting trajectory
records user audio and text, model responses, interaction timestamps, and token
log probabilities; CharDuplex is conditioned on user audio, whereas the LLM
judge evaluates the textual dialogue.

\paragraph{Reward computation.}
Each completed trajectory under character condition $c$ receives two
LLM-judged rewards and two deterministic timing rewards. The interaction reward
$r_{\mathrm{int}}\in[0,1]$ evaluates conversational coherence, responsiveness,
and relevance, while the character reward $r_{\mathrm{char}}\in[0,1]$ evaluates
adherence to the prescribed identity, style, and role constraints. Both are
computed from the textual dialogue; the character judge additionally receives
the character description, while the interaction judge receives timing
statistics. The corresponding judge prompts are provided in
Appendix~\ref{sec:app_rewardprompt}.

The timing rewards measure response initiation and yielding under interruption.
Specifically, $d_{\mathrm{resp}}$ denotes the delay from the end of a normal
user turn to the onset of the model response, and $d_{\mathrm{yield}}$ denotes
the delay from a user interruption to the model becoming silent. For successful
events, both rewards use
\begin{equation}
    g(d;D)=\max\!\left(0,1-\frac{d}{D}\right),
    \quad
    D_{\mathrm{resp}}=2\,\mathrm{s},
    \quad
    D_{\mathrm{yield}}=0.64\,\mathrm{s}.
    \label{eq:timing_reward}
\end{equation}
Failure to respond before the next user turn or the end of the rollout receives
$-1$, as does failure to yield during an interrupting user turn. Each timing
reward is averaged over its eligible events and is zero when no eligible event
occurs. The final trajectory reward is
\begin{equation}
    R(\tau,c)
    =
    r_{\mathrm{int}}
    +r_{\mathrm{char}}
    +r_{\mathrm{resp}}
    +r_{\mathrm{yield}}.
    \label{eq:fdgym_reward}
\end{equation}

\paragraph{Policy optimization.}
After full-duplex rollout and reward calculation, we use group relative policy optimization (GRPO) to train CharDuplex \citep{deepseekmath}. The detailed algorithm of GRPO is shown in Appendix \ref{sec:app_grpo}.
\begin{table}[t]
    \centering
    \small
    \setlength{\tabcolsep}{6pt}
    \caption{
        Results on SpeechRole-Eval.
        IA: Instruction Adherence;
        CC: Conversational Coherence;
        PeC: Personality Consistency;
        KC: Knowledge Consistency.
        Average is computed over the four metrics.
        Bold values indicate the best result within each model group.
    }
    \label{tab:speechrole_eval}
    \begin{threeparttable}
    \renewcommand{\arraystretch}{0.95}
    \begin{tabular}{lccccc}
        \toprule
        Model & IA $\uparrow$ & CC $\uparrow$ & PeC $\uparrow$ & KC $\uparrow$ & Avg. $\uparrow$ \\
        \midrule
        \multicolumn{6}{l}{\textit{Closed-source models}} \\
        Grok Voice Think Fast 2.0 \tnote{1}
            & 0.727 & 0.587 & 0.711 & 0.792 & 0.704 \\
        Gemini 3.1 Flash Live Preview \tnote{2}
            & \textbf{0.798} & \textbf{0.700} & \textbf{0.819} & \textbf{0.853} & \textbf{0.793} \\
        \midrule
        \multicolumn{6}{l}{\textit{Open-source models}} \\
        Moshi \citep{moshi}
            & 0.456 & 0.445 & 0.233 & 0.261 & 0.349 \\
        PersonaPlex \citep{personaplex}
            & 0.750 & 0.725 & 0.552 & 0.578 & 0.651 \\
        Freeze-Omni \citep{freezeomni}
            & 0.781 & 0.798 & 0.595 & 0.623 & 0.699 \\
        MiniCPM-o4.5 \citep{minicpm}
            & \textbf{0.815} & 0.747 & 0.655 & 0.727 & 0.736 \\
        CharDuplex-SFT
            & 0.807 & 0.800 & 0.729 & 0.770 & 0.777 \\
        CharDuplex
            & \textbf{0.815} & \textbf{0.805} & \textbf{0.738} & \textbf{0.781} & \textbf{0.785} \\
        \bottomrule
    \end{tabular}
    \begin{tablenotes}
        \footnotesize 
        \item[1] https://ai.google.dev/gemini-api/docs/models/gemini-3.1-flash-live-preview
    \end{tablenotes}
    \begin{tablenotes}
        \footnotesize 
        \item[2] https://x.ai/news/grok-voice-think-fast-2
    \end{tablenotes}
    \end{threeparttable}
\vspace{-10pt}
\end{table}

\section{Experimental Setup}
\label{sec:experimental_setup}

\subsection{Data and Training Details}
The pipeline in Section~\ref{sec:character_data} is applied to character descriptions from OmniCharacter and the SpeechRole training set \citep{omnicharacter,speechrole}. Qwen3.5-397B-A17B\footnote{https://huggingface.co/Qwen/Qwen3.5-397B-A17B}  serves as the generator, while GPT-5.6-Luna\footnote{https://developers.openai.com/api/docs/models/gpt-5.6-luna} serves as the verifier. Qwen3-TTS\footnote{https://huggingface.co/Qwen/Qwen3-TTS-12Hz-1.7B-CustomVoice} and Qwen3-ForcedAligner\footnote{https://huggingface.co/Qwen/Qwen3-ForcedAligner-0.6B} are used for speech synthesis and word-level alignment, respectively. The resulting corpus contains $70$ characters and $40{,}608$ dialogues, totaling $756.04$ hours of speech. Character descriptions used for training are disjoint from those used for evaluation.

CharDuplex is initialized from GLM-4-Voice \citep{glm4voice} and trained with general full-duplex data before character-conditioned SFT. The two stages use learning rates of $5\times10^{-4}$ and $5\times10^{-6}$, respectively. 
RL starts from CharDuplex-SFT. Qwen-235B\footnote{https://huggingface.co/Qwen/Qwen3-235B-A22B} is used for both user simulation and LLM-based reward computation, and Qwen3-TTS synthesizes user speech. Each policy iteration samples eight trajectories for each of four groups, yielding 32 trajectories in total. GRPO uses a learning rate of $2\times10^{-6}$. Detailed rollout, sampling, and optimization configurations are provided in Appendix~\ref{sec:app_trainingdetails}.

\subsection{Benchmarks and Evaluation}

\paragraph{Character-related capabilities.}
SpeechRole-Eval \citep{speechrole} assesses interaction quality and role-playing fidelity. Four metrics are reported: Instruction Adherence (IA), Conversational Coherence (CC), Personality Consistency (PeC), and Knowledge Consistency (KC). All four metrics are evaluated in the prescribed role-playing setting. IA assesses adherence to the role-playing instructions, including remaining in character and avoiding out-of-role explanations. CC measures logical consistency with the preceding dialogue, while PeC and KC assess consistency with the character's personality traits and established knowledge, respectively. Each evaluation dialogue is conditioned on its associated character description.

\paragraph{General speech intelligence and conversational behavior.}
General speech intelligence is evaluated on AlpacaEval, CommonEval, WildVoice, OpenBookQA, MMSU, SD-QA, BBH, and AdvBench from VoiceBench \citep{voicebench}, as well as LLaMA-Questions \citep{llamaq} and WebQuestions \citep{webquestions}. These datasets cover spoken question answering, reasoning, open-ended response quality, and safety. Full-Duplex-Bench \citep{fullduplex-bench} further evaluates turn-taking and interruption handling in full-duplex interaction. Throughout the experiments, \textbf{CharDuplex-SFT} refers to the character-conditioned SFT model, and \textbf{CharDuplex} refers to the final model after FDGym-based RL.
\section{Results}
\label{sec:results}
\subsection{Character-Related Capabilities}
\label{sec:character_results}
Table~\ref{tab:speechrole_eval} compares CharDuplex with representative open- and
closed-source full-duplex SDLMs on SpeechRole-Eval. CharDuplex reaches an average score of 0.785, improving over PersonaPlex by 0.134 absolute and outperforming it across all four dimensions. The largest gains occur in Personality Consistency (+0.186) and Knowledge Consistency (+0.203), indicating that the improvement extends beyond instruction following to the two dimensions most directly associated with maintaining character identity and background.
Among the evaluated open-source models, CharDuplex achieves the highest average
score, leads on Conversational Coherence, Personality Consistency, and Knowledge
Consistency, and ties for the best Instruction Adherence. These results indicate
that the improvement is not limited to following role-playing instructions, but
extends to maintaining the character's personality and knowledge across the
conversation.

Compared with the closed-source systems, CharDuplex remains competitive overall.
It exceeds Grok on four dimensions and Gemini on Instruction Adherence and Conversational
Coherence, while Gemini retains stronger Personality Consistency and Knowledge
Consistency. More importantly, CharDuplex-SFT already exhibits strong
character-consistent behavior before RL, supporting the effectiveness of the
character-conditioned data and SFT stage. FDGym subsequently improves all four
SpeechRole-Eval dimensions, showing that interactive post-training further
refines character consistency on top of an already strong supervised model.

\begin{table}[t]
    \centering
    \small
    \setlength{\tabcolsep}{3.5pt}
    \caption{
        General speech intelligence results. FD indicates full-duplex
        support. LlamaQ, WebQ, OBQA, AdvB., AlpacaE, ComE, and WildV
        denote LLaMA-Questions, WebQuestions, OpenBookQA, AdvBench,
        AlpacaEval, CommonEval, and WildVoice, respectively. Published baseline results are collected
        from prior works\citep{flair,steerduplex}. ``--'' denotes an unreported result.
        Bold marks the best score
        separately among full-duplex or non-full-duplex models.
    }
    \label{tab:speech_intelligence}
    \renewcommand{\arraystretch}{1.0}
    \resizebox{\linewidth}{!}{%
    \begin{tabular}{lc*{10}{c}}
        \toprule
        & & \multicolumn{6}{c}{Accuracy (\%)}$\uparrow$
          & Refusal (\%)$\uparrow$ & \multicolumn{3}{c}{GPT-Score (1--5)}$\uparrow$ \\
        \cmidrule(lr){3-8}\cmidrule(lr){9-9}\cmidrule(lr){10-12}
        Model & FD & LlamaQ & WebQ & OBQA & MMSU & SD-QA
              & BBH & AdvB. & AlpacaE & ComE & WildV \\
        \midrule
        GLM-4-Voice & \ding{55}
            & 65.7 & 37.0 & 53.4 & 39.8 & 37.0
            & 52.8 & 88.08 & 3.97 & 3.42 & 3.18 \\
        Qwen2-Audio & \ding{55}
            & 69.7 & \textbf{45.2} & 49.5 & 35.7 & 35.7
            & 54.7 & 96.73 & 3.74 & 3.43 & 3.01 \\
        Kimi-Audio & \ding{55}
            & 68.3 & 37.3 & \textbf{83.5} & \textbf{62.2} & \textbf{63.1}
            & \textbf{69.7} & \textbf{100.00} & \textbf{4.46} & 3.97 & \textbf{4.20} \\
        Baichuan-Audio & \ding{55}
            & \textbf{74.0} & 40.7 & 71.7 & 53.2 & 45.8
            & 54.8 & 99.42 & 4.41 & \textbf{4.08} & 3.92 \\
        \midrule
        Moshi & \ding{51}
            & 54.5 & 22.1 & 25.9 & 24.0 & 15.6
            & 47.4 & 44.23 & 2.01 & 1.60 & 1.30 \\
        Freeze-Omni & \ding{51}
            & 56.2 & 27.9 & 31.0 & 28.1 & \textbf{53.5}
            & 50.7 & \textbf{97.30} & \textbf{4.03} & \textbf{3.46} & \textbf{3.15} \\
        SALMONN-omni & \ding{51}
            & 73.6 & \textbf{43.7} & -- & 30.0 & --
            & -- & -- & 3.22 & -- & -- \\
        PersonaPlex & \ding{51}
            & 46.2 & 21.5 & 27.2 & 27.8 & 23.4
            & 47.9 & 12.50 & 2.92 & 2.01 & 2.05 \\
        SteerDuplex & \ding{51}
            & - & - & - & -
            & 26.88 & 49.9 & 99.3 & 2.28 & 2.21 & 1.74 \\
        \midrule
        CharDuplex-SFT & \ding{51}
            & 74.3 & 40.8 & \textbf{63.1} & \textbf{44.2} & 47.4
            & 50.0 & 95.8 & 3.63 & 3.09 & 2.30 \\
        CharDuplex & \ding{51}
            & \textbf{74.7} & 40.5 & 62.4 & 44.0 & 49.2
            & \textbf{51.2} & 95.6 & 3.63 & 3.09 & 2.29 \\
        \bottomrule
    \end{tabular}%
    }
\vspace{-10pt}
\end{table}

\subsection{Speech Intelligence}
\label{sec:general_results}
We further evaluate whether the improved character consistency of CharDuplex
comes at the expense of general speech intelligence. As shown in
Table~\ref{tab:speech_intelligence}, CharDuplex compares favorably with
established full-duplex models across the evaluated benchmarks and exceeds its
GLM-4-Voice foundation on all five factual and multiple-choice QA tasks.
These tasks do not rely on character descriptions, indicating that the resulting
model retains general knowledge and response capability beyond role-playing.

Comparing CharDuplex-SFT with CharDuplex, FDGym improves all four
SpeechRole-Eval dimensions while the general benchmark results remain broadly
stable. Across the ten general speech-intelligence metrics, eight change by no more than 1.11\% relative after FDGym, while the two larger changes are improvements on SD-QA (+3.80\%) and BBH (+2.40\%). This pattern indicates that the character-consistency gains from interactive post-training are obtained without a systematic loss of general speech capability.

\begin{table*}[t]
    \centering
    \small
    \setlength{\tabcolsep}{5.5pt}
    \caption{
        Results on Full-Duplex-Bench \citep{fullduplex-bench}.
        TOR denotes turn-over rate, ICC denotes interjectional conversational cue,
        and JSD denotes Jensen--Shannon divergence.
        $\uparrow$ and $\downarrow$ indicate that higher and lower values are better, respectively.
        Bold values indicate the best result among open-source models, including CharDuplex.
    }
    \label{tab:full_duplex_bench}
    \resizebox{\textwidth}{!}{
    \begin{threeparttable}
    \renewcommand{\arraystretch}{1.1}
    \begin{tabular}{lcccccccccc}
        \toprule
        \multirow{2}{*}{Model}
        & \multicolumn{2}{c}{Pause Handling}
        & \multicolumn{3}{c}{Backchannel}
        & \multicolumn{2}{c}{Smooth Turn Taking}
        & \multicolumn{3}{c}{User Interruption} \\
        \cmidrule(lr){2-3}
        \cmidrule(lr){4-6}
        \cmidrule(lr){7-8}
        \cmidrule(lr){9-11}
        & Synthetic TOR $\downarrow$
        & Candor TOR $\downarrow$
        & TOR $\downarrow$
        & ICC Freq $\uparrow$
        & JSD $\downarrow$
        & TOR $\uparrow$
        & Candor Latency $\downarrow$
        & TOR $\uparrow$
        & Synthetic GPT-4o $\uparrow$
        & Latency $\downarrow$ \\
        \midrule

        dGSLM
        & 0.934
        & 0.935
        & 0.691
        & 0.015
        & 0.934
        & \textbf{0.975}
        & 0.352
        & 0.917
        & 0.201
        & 2.531 \\

        Moshi
        & 0.985
        & 0.980
        & 1.000
        & 0.001
        & 0.957
        & 0.941
        & 0.265
        & \textbf{1.000}
        & 0.765
        & \textbf{0.257} \\

        Freeze-Omni
        & \textbf{0.642}
        & \textbf{0.481}
        & \textbf{0.636}
        & 0.001
        & 0.997
        & 0.336
        & 0.953
        & 0.867
        & 3.615
        & 1.409 \\

        {Gemini Live\tnote{1}}
        & 
        {0.255}
        & {0.310}
        & {0.091}
        & {0.012}
        & {0.896}
        & {0.655}
        & {1.301}
        & {0.891}
        & {3.376}
        & {1.183} \\

        \midrule

        CharDuplex-SFT
        & 0.934
        & 0.815
        & \textbf{0.636}
        & 0.054
        & 0.890
        & 0.840
        & \textbf{0.133}
        & 0.990
        & \textbf{4.081}
        & 0.558 \\

        CharDuplex
        & 0.927
        & 0.815
        & 0.709
        & \textbf{0.055}
        & \textbf{0.889}
        & 0.857
        & 0.136
        & 0.990
        & 4.071
        & 0.565 \\

        \bottomrule
    \end{tabular}
    \begin{tablenotes}
        \footnotesize 
        \item[1] https://ai.google.dev/gemini-api/docs/live
    \end{tablenotes}
    \end{threeparttable}
    }
\vspace{-10pt}
\end{table*}

\subsection{Full-duplex interaction capability}
We evaluate the interaction performance of CharDuplex on Full-Duplex-Bench \citep{fullduplex-bench}. As shown in Table~\ref{tab:full_duplex_bench}, CharDuplex exhibits
strong full-duplex interaction capabilities across multiple dimensions.
In particular, it performs strongly on backchannel behavior and user
interruption handling, achieving competitive backchannel statistics and
a high interruption-quality score while maintaining a high interruption
TOR. The results indicate that the character-oriented training pipeline
does not come at the cost of the interaction behaviors required for
full-duplex conversation.

The comparison between CharDuplex-SFT and CharDuplex further shows that
FDGym-based RL largely preserves the full-duplex interaction profile
established during supervised training. Nine of the ten Full-Duplex-Bench metrics change by at most 2.26\% relative after FDGym. Together, these results suggest that interactive post-training
can strengthen character-consistent response generation without degrading the model's general intelligence and full-duplex interaction
capabilities. A consolidated comparison of CharDuplex-SFT and CharDuplex across character
consistency, general speech intelligence, and full-duplex interaction is
provided in Appendix~\ref{sec:app_sft_rl_comparison}. Further ablation studies and analysis for FDGym and the data-construction pipeline are provided in
Appendix~\ref{sec:app_ablation}.
\section{Conclusion}

This work presents CharDuplex, a character-driven full-duplex SDLM that combines character-consistent response generation
with general speech intelligence and full-duplex interaction capabilities.
CharDuplex integrates an always-on dual-stream architecture, automated
character-conditioned data construction, and FDGym, which enables
response-dependent multi-turn RL with a simulated user. CharDuplex achieves
the highest average SpeechRole-Eval score among the evaluated open-source
models while remaining competitive with closed-source systems. FDGym further
improves all four character-related metrics, while general speech intelligence
and full-duplex interaction performance remain broadly stable. These results
show that character consistency can be strengthened without sacrificing the
general capabilities required for full-duplex spoken interaction.

\clearpage
\section*{AI Use Statement}

Generative AI tools were used to assist with language polishing, grammatical correction, and improving the clarity of the manuscript. AI models also served as components of the research methodology, including character-conditioned dialogue generation and verification of generated data, user simulation and reward computation during reinforcement learning, and automated evaluation and metric computation where applicable. The specific models, prompts, and procedures used for these purposes are described in the corresponding sections of the paper. All research ideas, methodological design, experiments, analysis, and conclusions were developed and verified by the authors. The authors reviewed all AI-assisted content and take full responsibility for the manuscript and reported results.

\bibliography{iclr2027_conference}
\bibliographystyle{iclr2027_conference}

\clearpage
\appendix
\begin{figure}[htbp]
\begin{tcolorbox}[
    colback=gray!5!white,
    colframe=gray!75!black,
    title=Prompt Template for Character-Conditioned Dialogue Generation,
    arc=2mm,
    boxrule=0.5pt
]
\small

\textbf{[Task]}\\
Create one realistic and useful multi-turn spoken conversation for full-duplex
dialogue training. Maintain one causally connected situation and one primary
conversational goal. Every turn should respond naturally to the preceding
dialogue and provide a plausible reason for the next turn.\vspace{0.5em}

\textbf{[Character and Scenario]}\\
Character / role description: \{Character\_Description\}\\
Scenario: \{Scenario\}\\
Conversational goal: \{Goal\}\vspace{0.5em}

\textbf{[Character Grounding]}\\
Construct a private \texttt{system\_instruction} containing the character's
relevant personality, background, and role-specific knowledge. Any private,
current, or scenario-specific fact later stated by the character must already
be supported by this instruction or by information previously provided by
the user. User-specific facts must originate from the user.\vspace{0.5em}

\textbf{[Dialogue Requirements]}\\
- Keep the conversation internally coherent and grounded in the available information.\\
- Do not invent unsupported prices, schedules, policies, actions, observations,
or other private/current facts.\\
- If required information is unavailable, ask for it or express uncertainty.\\
- Conclusions and calculations must follow from information already established
in the dialogue.\\
- Do not include interaction-event annotations, timing markers, or stage directions
in the dialogue text.\vspace{0.5em}

\textbf{[Output Format]}\\
Return only one JSON object:
\begin{quote}\ttfamily
\{"system\_instruction": "...",\\
\hspace*{0.8em}"turns": [\\
\hspace*{1.6em}\{"speaker": "s1", "text": "..."\},\\
\hspace*{1.6em}\{"speaker": "s0", "text": "..."\}, ... ]\}
\end{quote}

Use only \texttt{s0} for the character and \texttt{s1} for the user.
Keep speaker turns alternating and end with \texttt{s0}.
\end{tcolorbox}
\caption{
Prompt template used by the generator to construct character-conditioned
multi-turn dialogues. The character description and sampled scenario define
the conversational context, while explicit grounding constraints require
role-specific claims to be supported by the character background or preceding
dialogue.
}
\label{fig:generation_prompt}
\end{figure}
\begin{figure}[htbp]
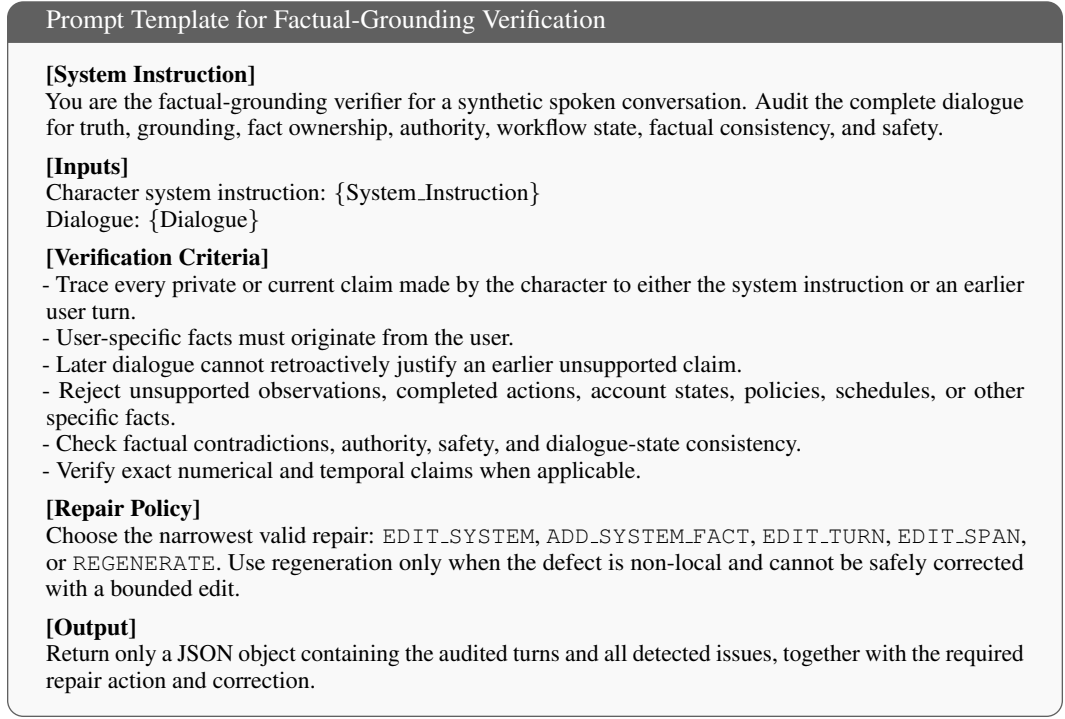

\begin{tcolorbox}[
    colback=gray!5!white,
    colframe=gray!75!black,
    title=Prompt Template for Factual-Grounding Verification,
    arc=2mm,
    boxrule=0.5pt
]
\small

\textbf{[System Instruction]}\\
You are the factual-grounding verifier for a synthetic spoken conversation.
Audit the complete dialogue for truth, grounding, fact ownership, authority,
workflow state, factual consistency, and safety.\vspace{0.5em}

\textbf{[Inputs]}\\
Character system instruction: \{System\_Instruction\}\\
Dialogue: \{Dialogue\}\vspace{0.5em}

\textbf{[Verification Criteria]}\\
- Trace every private or current claim made by the character to either the
system instruction or an earlier user turn.\\
- User-specific facts must originate from the user.\\
- Later dialogue cannot retroactively justify an earlier unsupported claim.\\
- Reject unsupported observations, completed actions, account states,
policies, schedules, or other specific facts.\\
- Check factual contradictions, authority, safety, and dialogue-state
consistency.\\
- Verify exact numerical and temporal claims when applicable.\vspace{0.5em}

\textbf{[Repair Policy]}\\
Choose the narrowest valid repair:
\texttt{EDIT\_SYSTEM}, \texttt{ADD\_SYSTEM\_FACT},
\texttt{EDIT\_TURN}, \texttt{EDIT\_SPAN}, or \texttt{REGENERATE}.
Use regeneration only when the defect is non-local and cannot be safely
corrected with a bounded edit.\vspace{0.5em}

\textbf{[Output]}\\
Return only a JSON object containing the audited turns and all detected issues,
together with the required repair action and correction.
\end{tcolorbox}
\caption{
Prompt template used by the factual-grounding verifier. The verifier checks
whether character-specific claims are supported by the prescribed background
or previously established dialogue context and returns localized repairs when
possible.
}
\label{fig:factual_prompt}
\end{figure}

\begin{figure}[t]
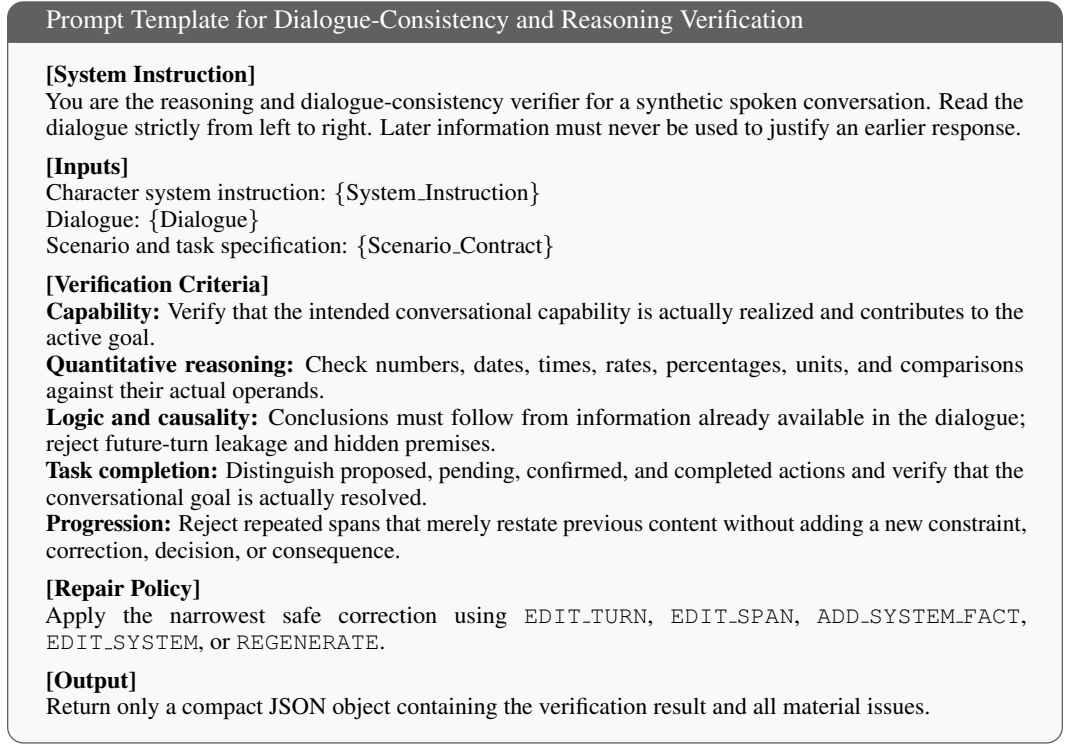

\begin{tcolorbox}[
    colback=gray!5!white,
    colframe=gray!75!black,
    title=Prompt Template for Dialogue-Consistency and Reasoning Verification,
    arc=2mm,
    boxrule=0.5pt
]
\small

\textbf{[System Instruction]}\\
You are the reasoning and dialogue-consistency verifier for a synthetic spoken
conversation. Read the dialogue strictly from left to right. Later information
must never be used to justify an earlier response.\vspace{0.5em}

\textbf{[Inputs]}\\
Character system instruction: \{System\_Instruction\}\\
Dialogue: \{Dialogue\}\\
Scenario and task specification: \{Scenario\_Contract\}\vspace{0.5em}

\textbf{[Verification Criteria]}\\
\textbf{Capability:} Verify that the intended conversational capability is
actually realized and contributes to the active goal.\\
\textbf{Quantitative reasoning:} Check numbers, dates, times, rates,
percentages, units, and comparisons against their actual operands.\\
\textbf{Logic and causality:} Conclusions must follow from information already
available in the dialogue; reject future-turn leakage and hidden premises.\\
\textbf{Task completion:} Distinguish proposed, pending, confirmed, and
completed actions and verify that the conversational goal is actually resolved.\\
\textbf{Progression:} Reject repeated spans that merely restate previous
content without adding a new constraint, correction, decision, or consequence.
\vspace{0.5em}

\textbf{[Repair Policy]}\\
Apply the narrowest safe correction using
\texttt{EDIT\_TURN}, \texttt{EDIT\_SPAN},
\texttt{ADD\_SYSTEM\_FACT}, \texttt{EDIT\_SYSTEM},
or \texttt{REGENERATE}.\vspace{0.5em}

\textbf{[Output]}\\
Return only a compact JSON object containing the verification result and
all material issues.
\end{tcolorbox}
\caption{
Prompt template used by the second semantic verifier to assess cross-turn
reasoning, chronology, task completion, and dialogue progression. It complements
the factual-grounding verifier by focusing on dependencies and consistency
across the multi-turn interaction.
}
\label{fig:consistency_prompt}
\end{figure}
\begin{figure}[t]
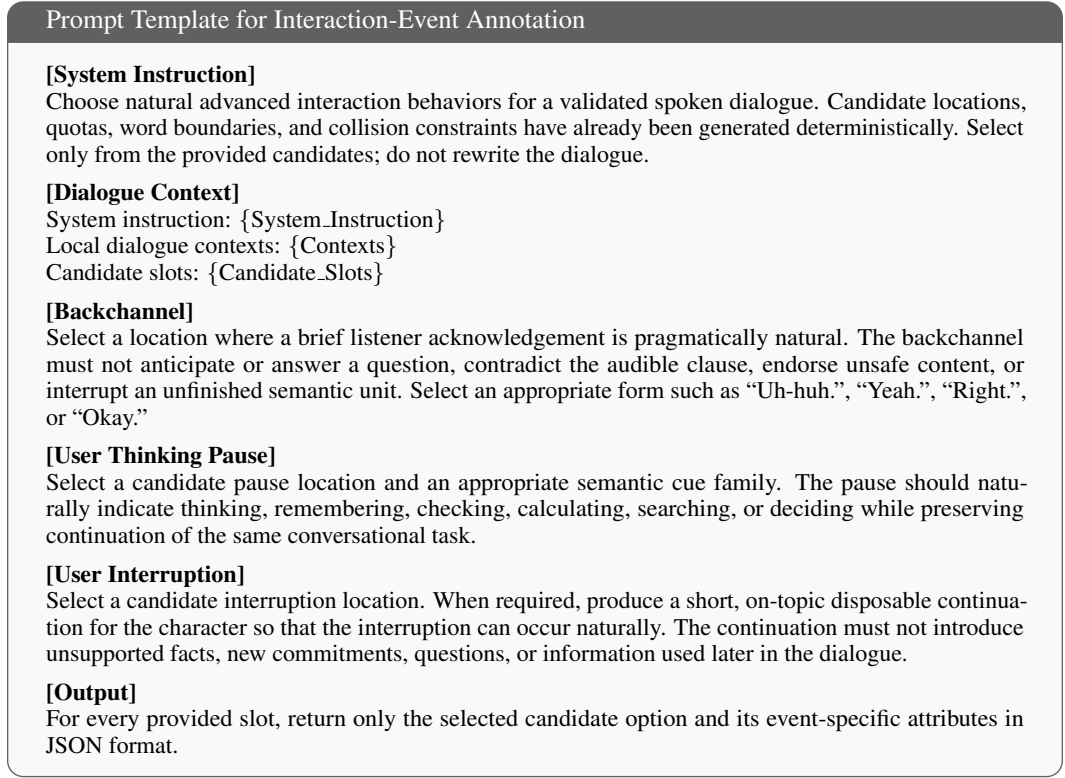

\begin{tcolorbox}[
    colback=gray!5!white,
    colframe=gray!75!black,
    title=Prompt Template for Interaction-Event Annotation,
    arc=2mm,
    boxrule=0.5pt
]
\small

\textbf{[System Instruction]}\\
Choose natural advanced interaction behaviors for a validated spoken dialogue.
Candidate locations, quotas, word boundaries, and collision constraints have
already been generated deterministically. Select only from the provided
candidates; do not rewrite the dialogue.\vspace{0.5em}

\textbf{[Dialogue Context]}\\
System instruction: \{System\_Instruction\}\\
Local dialogue contexts: \{Contexts\}\\
Candidate slots: \{Candidate\_Slots\}\vspace{0.5em}

\textbf{[Backchannel]}\\
Select a location where a brief listener acknowledgement is pragmatically
natural. The backchannel must not anticipate or answer a question, contradict
the audible clause, endorse unsafe content, or interrupt an unfinished
semantic unit. Select an appropriate form such as
``Uh-huh.'', ``Yeah.'', ``Right.'', or ``Okay.''\vspace{0.5em}

\textbf{[User Thinking Pause]}\\
Select a candidate pause location and an appropriate semantic cue family.
The pause should naturally indicate thinking, remembering, checking,
calculating, searching, or deciding while preserving continuation of the
same conversational task.\vspace{0.5em}

\textbf{[User Interruption]}\\
Select a candidate interruption location. When required, produce a short,
on-topic disposable continuation for the character so that the interruption
can occur naturally. The continuation must not introduce unsupported facts,
new commitments, questions, or information used later in the dialogue.
\vspace{0.5em}

\textbf{[Output]}\\
For every provided slot, return only the selected candidate option and its
event-specific attributes in JSON format.
\end{tcolorbox}
\caption{
Prompt template used to annotate full-duplex interaction events. Candidate
locations are generated deterministically, while the generator selects
contextually appropriate backchannels, user thinking pauses, and interruptions
without modifying the verified dialogue.
}
\label{fig:annotation_prompt}
\end{figure}
\begin{figure}[t]
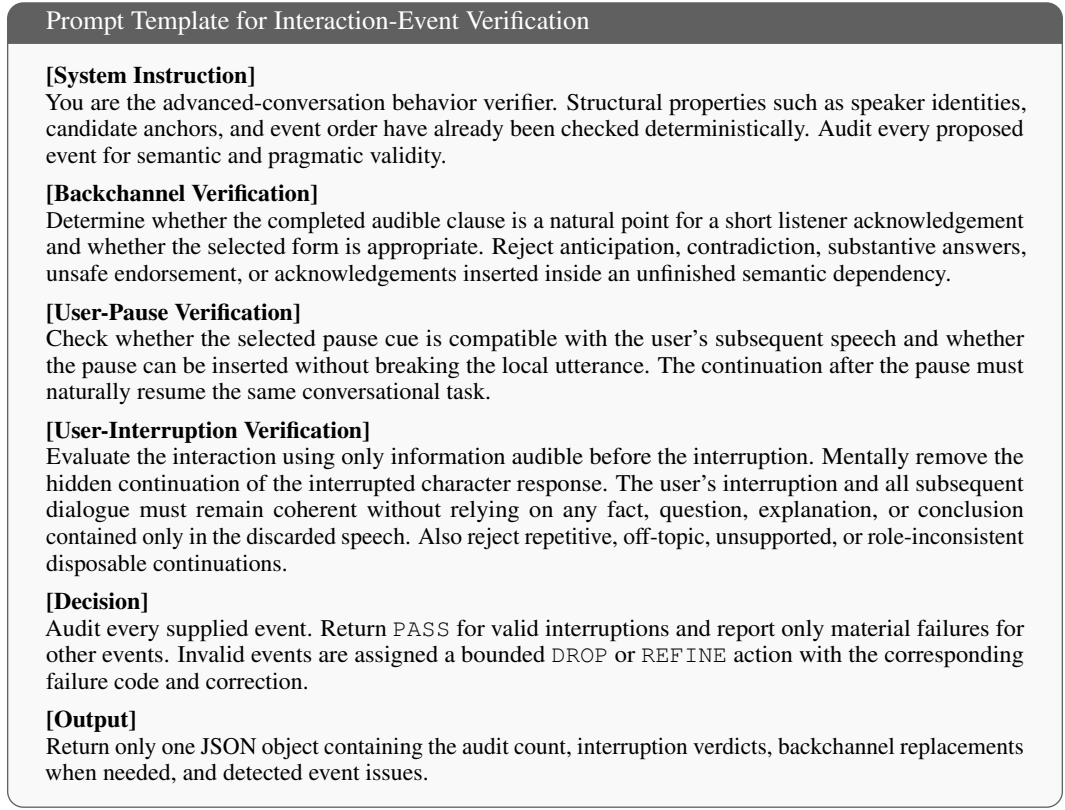

\begin{tcolorbox}[
    colback=gray!5!white,
    colframe=gray!75!black,
    title=Prompt Template for Interaction-Event Verification,
    arc=2mm,
    boxrule=0.5pt
]
\small

\textbf{[System Instruction]}\\
You are the advanced-conversation behavior verifier. Structural properties
such as speaker identities, candidate anchors, and event order have already
been checked deterministically. Audit every proposed event for semantic and
pragmatic validity.\vspace{0.5em}

\textbf{[Backchannel Verification]}\\
Determine whether the completed audible clause is a natural point for a short
listener acknowledgement and whether the selected form is appropriate.
Reject anticipation, contradiction, substantive answers, unsafe endorsement,
or acknowledgements inserted inside an unfinished semantic dependency.
\vspace{0.5em}

\textbf{[User-Pause Verification]}\\
Check whether the selected pause cue is compatible with the user's subsequent
speech and whether the pause can be inserted without breaking the local
utterance. The continuation after the pause must naturally resume the same
conversational task.\vspace{0.5em}

\textbf{[User-Interruption Verification]}\\
Evaluate the interaction using only information audible before the interruption.
Mentally remove the hidden continuation of the interrupted character response.
The user's interruption and all subsequent dialogue must remain coherent
without relying on any fact, question, explanation, or conclusion contained
only in the discarded speech. Also reject repetitive, off-topic, unsupported,
or role-inconsistent disposable continuations.\vspace{0.5em}

\textbf{[Decision]}\\
Audit every supplied event. Return \texttt{PASS} for valid interruptions and
report only material failures for other events. Invalid events are assigned a
bounded \texttt{DROP} or \texttt{REFINE} action with the corresponding failure
code and correction.\vspace{0.5em}

\textbf{[Output]}\\
Return only one JSON object containing the audit count, interruption verdicts,
backchannel replacements when needed, and detected event issues.
\end{tcolorbox}
\caption{
Prompt template used by the event-specific verifier. Backchannels and pauses
are checked for local pragmatic compatibility, while interruptions are
additionally validated under the information actually audible before the
interruption, preventing subsequent dialogue from relying on discarded speech.
}
\label{fig:event_verification_prompt}
\end{figure}

\begin{figure}[t]
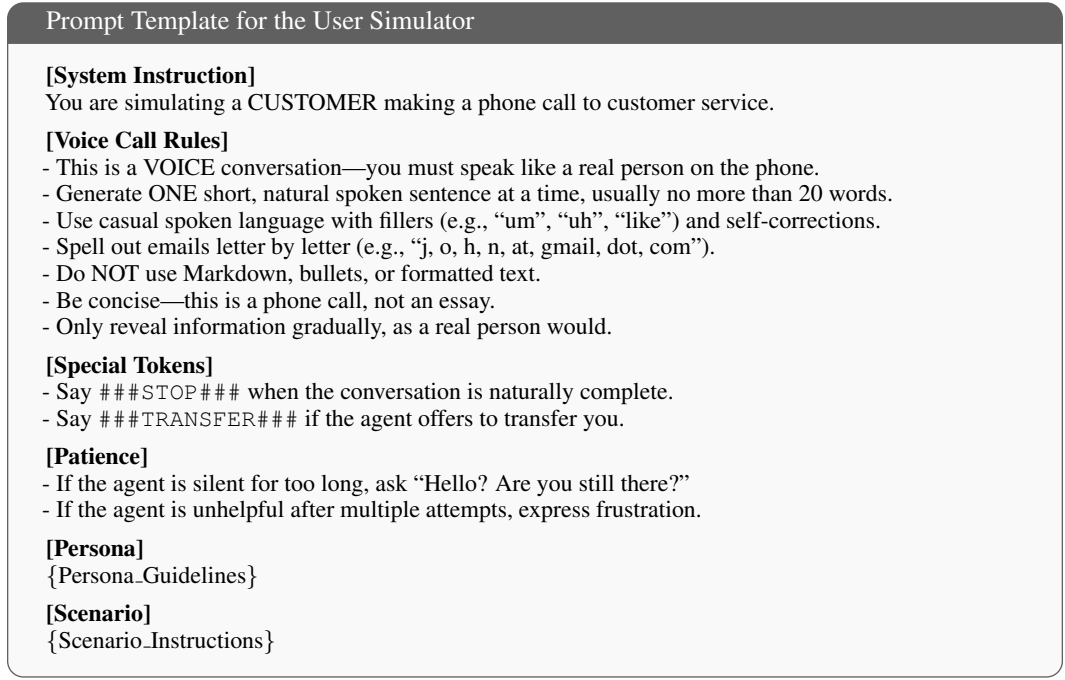

\begin{tcolorbox}[
    colback=gray!5!white,
    colframe=gray!75!black,
    title=Prompt Template for the User Simulator,
    arc=2mm,
    boxrule=0.5pt
]
\small

\textbf{[System Instruction]}\\
You are simulating a CUSTOMER making a phone call to customer service.
\vspace{0.5em}

\textbf{[Voice Call Rules]}\\
- This is a VOICE conversation---you must speak like a real person on the phone.\\
- Generate ONE short, natural spoken sentence at a time, usually no more than
20 words.\\
- Use casual spoken language with fillers (e.g., ``um'', ``uh'', ``like'')
and self-corrections.\\
- Spell out emails letter by letter (e.g., ``j, o, h, n, at, gmail, dot, com'').\\
- Do NOT use Markdown, bullets, or formatted text.\\
- Be concise---this is a phone call, not an essay.\\
- Only reveal information gradually, as a real person would.
\vspace{0.5em}

\textbf{[Special Tokens]}\\
- Say \texttt{\#\#\#STOP\#\#\#} when the conversation is naturally complete.\\
- Say \texttt{\#\#\#TRANSFER\#\#\#} if the agent offers to transfer you.
\vspace{0.5em}

\textbf{[Patience]}\\
- If the agent is silent for too long, ask ``Hello? Are you still there?''\\
- If the agent is unhelpful after multiple attempts, express frustration.
\vspace{0.5em}

\textbf{[Persona]}\\
\{Persona\_Guidelines\}
\vspace{0.5em}

\textbf{[Scenario]}\\
\{Scenario\_Instructions\}

\end{tcolorbox}

\caption{
Prompt template used by the LLM-based user simulator in FDGym.
The simulator produces short, natural spoken user turns conditioned on the
provided persona and dialogue scenario, revealing information progressively
as the interaction evolves.
}
\label{fig:user_simulator_prompt}
\end{figure}

\begin{figure}[t]
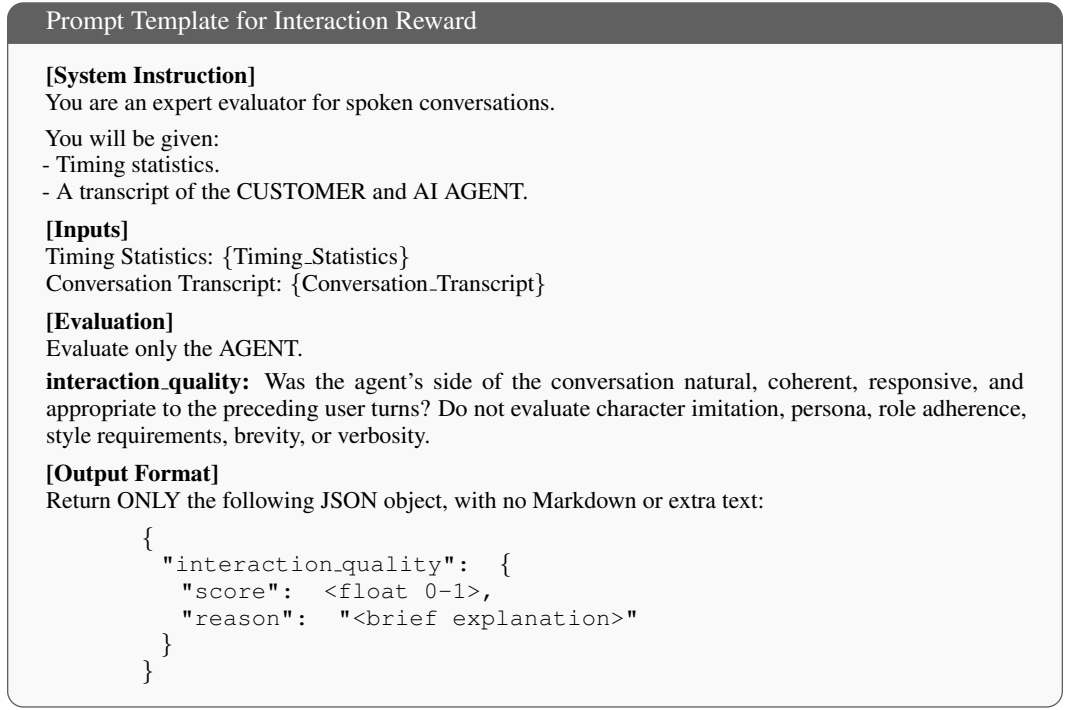

\begin{tcolorbox}[
    colback=gray!5!white,
    colframe=gray!75!black,
    title=Prompt Template for Interaction Reward,
    arc=2mm,
    boxrule=0.5pt
]
\small

\textbf{[System Instruction]}\\
You are an expert evaluator for spoken conversations.\vspace{0.4em}

You will be given:\\
- Timing statistics.\\
- A transcript of the CUSTOMER and AI AGENT.\vspace{0.5em}

\textbf{[Inputs]}\\
Timing Statistics: \{Timing\_Statistics\}\\
Conversation Transcript: \{Conversation\_Transcript\}\vspace{0.5em}

\textbf{[Evaluation]}\\
Evaluate only the AGENT.\vspace{0.3em}

\textbf{interaction\_quality:}
Was the agent's side of the conversation natural, coherent, responsive, and
appropriate to the preceding user turns? Do not evaluate character imitation,
persona, role adherence, style requirements, brevity, or verbosity.\vspace{0.5em}

\textbf{[Output Format]}\\
Return ONLY the following JSON object, with no Markdown or extra text:
\begin{quote}\ttfamily
\{\\
\hspace*{0.8em}"interaction\_quality": \{\\
\hspace*{1.6em}"score": <float 0-1>,\\
\hspace*{1.6em}"reason": "<brief explanation>"\\
\hspace*{0.8em}\}\\
\}
\end{quote}
\end{tcolorbox}

\caption{
Prompt template used to compute the interaction reward
$r_{\mathrm{int}}$. The judge evaluates the model's conversational coherence,
responsiveness, and appropriateness from the rollout transcript and timing
statistics, independently of character consistency.
}
\label{fig:interaction_reward_prompt}
\end{figure}

\begin{figure}[t]
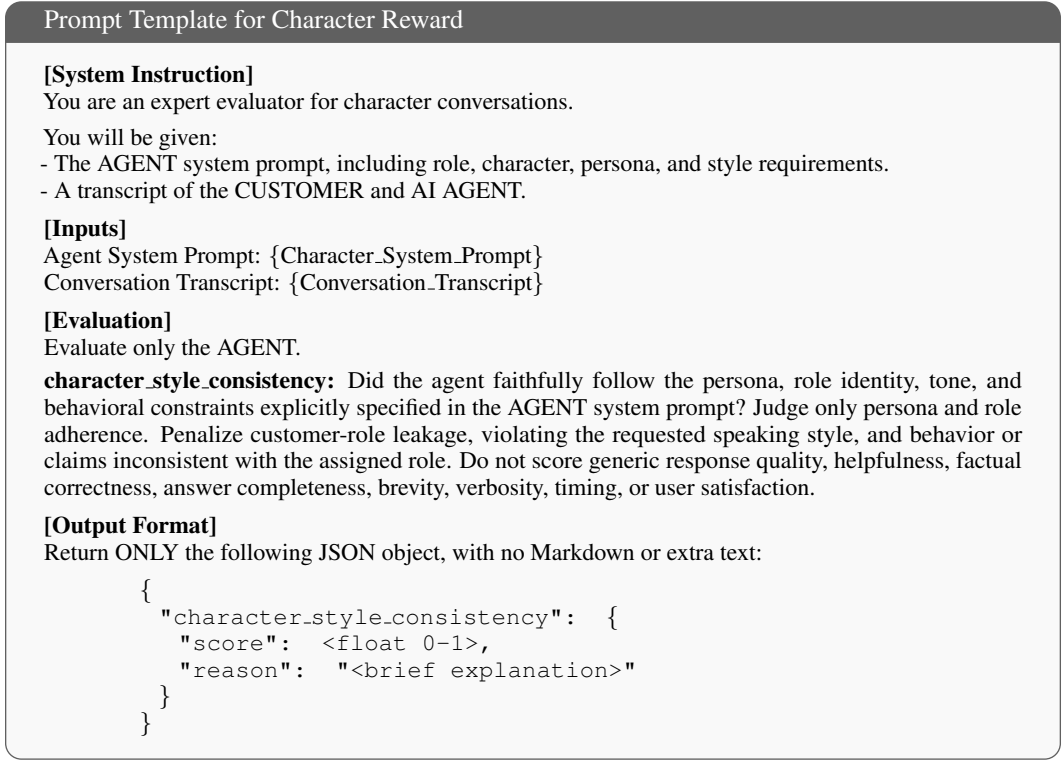

\begin{tcolorbox}[
    colback=gray!5!white,
    colframe=gray!75!black,
    title=Prompt Template for Character Reward,
    arc=2mm,
    boxrule=0.5pt
]
\small

\textbf{[System Instruction]}\\
You are an expert evaluator for character conversations.\vspace{0.4em}

You will be given:\\
- The AGENT system prompt, including role, character, persona, and style
requirements.\\
- A transcript of the CUSTOMER and AI AGENT.\vspace{0.5em}

\textbf{[Inputs]}\\
Agent System Prompt: \{Character\_System\_Prompt\}\\
Conversation Transcript: \{Conversation\_Transcript\}\vspace{0.5em}

\textbf{[Evaluation]}\\
Evaluate only the AGENT.\vspace{0.3em}

\textbf{character\_style\_consistency:}
Did the agent faithfully follow the persona, role identity, tone, and behavioral
constraints explicitly specified in the AGENT system prompt? Judge only persona
and role adherence. Penalize customer-role leakage, violating the requested
speaking style, and behavior or claims inconsistent with the assigned role.
Do not score generic response quality, helpfulness, factual correctness, answer
completeness, brevity, verbosity, timing, or user satisfaction.\vspace{0.5em}

\textbf{[Output Format]}\\
Return ONLY the following JSON object, with no Markdown or extra text:
\begin{quote}\ttfamily
\{\\
\hspace*{0.8em}"character\_style\_consistency": \{\\
\hspace*{1.6em}"score": <float 0-1>,\\
\hspace*{1.6em}"reason": "<brief explanation>"\\
\hspace*{0.8em}\}\\
\}
\end{quote}
\end{tcolorbox}

\caption{
Prompt template used to compute the character reward
$r_{\mathrm{char}}$. The judge evaluates whether the model's responses remain
consistent with the prescribed identity, persona, tone, and behavioral
constraints, independently of generic response quality and interaction timing.
}
\label{fig:character_reward_prompt}
\end{figure}

\section{Character-conditioned data construction}
\label{sec:app_char_data_generation}
We propose an \textit{automated generator--verifier pipeline} for textual dialogue generation that alternates between generation and verification. A \textit{generator} produces dialogue text and event annotations, while a \textit{verifier} checks and corrects these outputs using task-specific prompts. The pipeline comprises the following three steps.

\textit{(i) Dialogue generation.}
The generator receives a character description and a sampled conversational scenario. The description specifies the character's personality and background; the scenario supplies a topic and conversational goal. The generation prompt requires replies to reflect the prescribed character and respond to the preceding dialogue, which is shown in Figure \ref{fig:generation_prompt}. This conditioning associates character traits with responses in specific conversational situations. The character description is retained as the system prompt and excluded from the text passed to TTS.

\textit{(ii) Content verification and correction.}
The verifier reviews the generated dialogue using two separate prompts. The first checks factual support: role-specific claims must be supported by the character description or preceding dialogue, and facts about the user must not be attributed to the character. The corresponding prompt is shown in Figure \ref{fig:factual_prompt}. The second checks dialogue consistency: conclusions must follow from information already available, and subsequent replies must respect earlier statements and corrections, shown in Figure \ref{fig:consistency_prompt}. The verifier identifies erroneous turns and specifies corrections. An isolated error receives a turn-level edit; if the edit changes information used in subsequent replies, those replies are revised together. The verifier then checks the corrected dialogue again for unresolved errors or contradictions introduced by the edits. Further errors that invalidate the scenario or affect much of the conversation trigger regeneration of the whole conversation instead of local editing.

\textit{(iii) Interaction-event annotation.}
Once the dialogue passes verification, a rule-based procedure identifies candidate locations for backchannels, user thinking pauses, and interruptions. The generator LLM now acts as the event annotator, selecting suitable locations from these candidates according to the dialogue context. The prompt for annotation is shown in Figure \ref{fig:annotation_prompt}. The verifier checks the proposed events with an event-specific prompt, and incorrect annotations are revised. The prompt for the annotation verifier is shown in Figure \ref{fig:event_verification_prompt}. Each annotation records the speakers and text boundaries needed for Stage 3. Event annotations remain separate from the dialogue text, so the audio construction procedure controls pauses and overlaps without passing timing instructions to TTS.

\section{Data details}
\label{sec:app_datadetails}
\subsection{General Full-Duplex SFT Data}
\label{sec:app_general_sft_data}

For general full-duplex training, we use the same corpus and data construction
setup as \citep{flair, nvidialink}. The corpus contains three components:
approximately 530K hours of speech-continuation data, 70K hours of
instruction-following QA data, and 20K hours of ASR-QA data.

The speech-continuation data are constructed by converting continuous text
passages into alternating user--assistant turns and synthesizing the two
speakers separately. The instruction-following QA data contain both single-turn
and multi-turn conversations generated from diverse textual contexts, including
Wikipedia passages\footnote{
https://huggingface.co/datasets/wikimedia/wikipedia}, while the ASR-QA data are built from real speech in NeMo
ASRSET by generating questions grounded in the corresponding transcriptions.
Across the synthetic subsets, text generation uses GPT-OSS-120B\footnote{https://huggingface.co/openai/gpt-oss-120b},
Qwen2.5-72B-Instruct\footnote{
https://huggingface.co/Qwen/Qwen2.5-72B-Instruct}, and Llama-3.1-70B-Instruct\footnote{
https://huggingface.co/meta-llama/Llama-3.1-70B-Instruct}, and speech synthesis uses
Chatterbox\footnote{
https://github.com/resemble-ai/chatterbox}, Magpie-TTS\footnote{https://docs.nvidia.com/nemo-framework/user-guide/latest/speech\_ai/magpietts.html}, and MoonCast\footnote{https://github.com/jzq2000/MoonCast}. Voice references are sampled from
LibriTTS \citep{zen2019libritts}, YODAS \citep{li2023yodas}, and HiFi-TTS \citep{bakhturina2021hi}, yielding a prompt pool with over 100K segments
from more than 20K speakers.

During training, SpecAugment is applied to user speech features, and background
noise from Freesound \citep{freesound} and MUSAN \citep{musan} is mixed into the user stream for additional
acoustic diversity. Further details of the original construction procedure are
provided in \citet{flair}.

\paragraph{Character-conditioned data.}
The pipeline in Section~\ref{sec:character_data} is applied to character descriptions from OmniCharacter and the SpeechRole training set \citep{omnicharacter,speechrole}.Qwen3.5-397B-A17B\footnote{https://huggingface.co/Qwen/Qwen3.5-397B-A17B} serves as the generator for dialogue generation and event annotation, while GPT-5.6-Luna\footnote{https://developers.openai.com/api/docs/models/gpt-5.6-luna} serves as the verifier for content verification, correction, and event validation. Qwen3-TTS\footnote{https://huggingface.co/Qwen/Qwen3-TTS-12Hz-1.7B-CustomVoice} and Qwen3-ForcedAligner\footnote{https://huggingface.co/Qwen/Qwen3-ForcedAligner-0.6B} are used for speech synthesis and word-level alignment, respectively. The resulting corpus contains $70$ characters and $40{,}608$ dialogues, totaling $756.043$ hours of speech. We ensure that the character descriptions used for training are disjoint from those used for evaluation.

\section{Training Details}
\label{sec:app_trainingdetails}
\paragraph{Supervised training.}
CharDuplex is initialized from GLM-4-Voice \citep{glm4voice} and trained using NeMo Toolkit \citep{nemo}. The general training corpus comprises 530K hours of speech continuation data, 70K hours of instruction-following QA data, and 20K hours of ASR-QA data \citep{flair}. Character-conditioned SFT uses the additional corpus described above. Both supervised stages use AdamW with $(\beta_1,\beta_2)=(0.9,0.98)$ and zero weight decay. An inverse-square-root learning-rate schedule is applied, with a learning rate of $5\times10^{-4}$ and 2,500 warm-up steps for general full-duplex training, and $5\times10^{-6}$ and $1{,}000$ warm-up stepsfor character-conditioned SFT. The model is trained on 32 A100 GPUs. SpecAugment is applied to user speech features \citep{speechbrain}, and background noise from Freesound \citep{freesound} and MUSAN \citep{musan} is added to user audio with probability 0.5, with the signal-to-noise ratio sampled uniformly between 0 and 60\,dB. 

\paragraph{Interactive RL.}
RL starts from the character-conditioned SFT checkpoint. Qwen-235B is used for both user simulation and LLM-based reward computation, and Qwen3-TTS synthesizes user speech. Each policy iteration samples eight trajectories for each of four initial dialogue conditions, yielding 32 trajectories. Trajectories are limited to 260 ticks of 160\,ms, corresponding to 41.6\,s of simulated dialogue. Response sampling uses a temperature of 1.0 and top-$p$ of 0.9. GRPO uses a learning rate of $2\times10^{-6}$ and a clipping threshold of 0.2. Each collected batch is processed once with one trajectory per GPU on eight A100 GPUs, giving four optimizer steps per policy iteration. The RL budget is set to 500 policy iterations. During the rollout, only the portion that would have become
audible by the current tick is exposed to the simulator, using a speaking-rate estimate of 3.4 words per second derived from the SFT datasets.

\section{Prompts used in FDGym}
\label{sec:app_rewardprompt}
FDGym uses an LLM-based user simulator to generate response-dependent user
speech during each rollout. The simulator generates subsequent user utterances
conditioned on the dialogue history and the progressively available CharDuplex
response. Its prompt template is shown in
Figure~\ref{fig:user_simulator_prompt}.

FDGym uses two independent LLM-based judges to compute the interaction and
character reward components described in Section~\ref{sec:interactive_rl}. The interaction
judge receives timing statistics together with the textual dialogue, whereas
the character judge additionally conditions on the character system prompt.
Figures~\ref{fig:interaction_reward_prompt} and
\ref{fig:character_reward_prompt} show the corresponding prompt templates used
during RL. Dynamic rollout-dependent inputs are represented by placeholders.

\section{GRPO for CharDuplex}
\label{sec:app_grpo}
We optimize CharDuplex using group relative policy optimization (GRPO) \citep{deepseekmath}. Let $\pi_{\theta}(y\mid s)$ denote the response-token distribution of CharDuplex with trainable parameters $\theta$, conditioned on dialogue context $s$. At each iteration, we collect $G$ trajectories $\{\tau_i\}_{i=1}^{G}$ under the same character prompt and initial user utterance using a parameter snapshot $\theta_{\mathrm{old}}$. This rollout policy remains fixed while the collected trajectories are used to update $\theta$.

For trajectory reward $R_i=R(\tau_i,c)$, we compute the group-relative advantage
\begin{equation}
    A_i
    =
    \operatorname{clip}\!\left(
        \frac{R_i-\bar{R}}
             {\max(\sigma_R,\epsilon_{\mathrm{num}})},
        -A_{\max},A_{\max}
    \right),
    \label{eq:fdgym_advantage}
\end{equation}
where $\bar{R}=\frac{1}{G}\sum_{j=1}^{G}R_j$ and
$\sigma_R=\left[\frac{1}{G}\sum_{j=1}^{G}(R_j-\bar{R})^2\right]^{1/2}$
are the group reward mean and population standard deviation.
The positive constant $\epsilon_{\mathrm{num}}$ prevents division by zero,
and $A_{\max}$ bounds the advantage magnitude.

Let $y_{i,t}$ be the sampled response token at position $t$ in trajectory $i$.
Its context $s_{i,t}$ contains the character prompt, available user speech,
and preceding model outputs. We recompute its probability under the current
policy and form the ratio
\begin{equation}
    \rho_{i,t}(\theta)
    =
    \frac{
        \pi_{\theta}(y_{i,t}\mid s_{i,t})
    }{
        \pi_{\theta_{\mathrm{old}}}(y_{i,t}\mid s_{i,t})
    }.
    \label{eq:fdgym_policy_ratio}
\end{equation}
The clipped policy loss is computed over the complete response stream
of each sampled trajectory:
\begin{equation}
\begin{aligned}
    \mathcal{L}_{\mathrm{policy}}
    =
    -\frac{1}{N}
    \sum_{i=1}^{G}\sum_{t=1}^{T_i}
    \min\!\Bigl(
        &\rho_{i,t}(\theta)A_i,\\
        &\operatorname{clip}\!\left(
            \rho_{i,t}(\theta),1-\varepsilon,1+\varepsilon
        \right)A_i
    \Bigr),
\end{aligned}
\label{eq:fdgym_policy_loss}
\end{equation}
where $T_i$ is the number of response tokens generated in trajectory
$\tau_i$, $N=\sum_{i=1}^{G}T_i$ is the total number of generated response
tokens, and $\varepsilon>0$ controls probability-ratio clipping. After one optimization pass over
the collected trajectories, the updated model is used for the next
rollout iteration.

\section{Comparison Before and After Interactive Post-Training}
\label{sec:app_sft_rl_comparison}

To further examine the effect of FDGym, Table~\ref{tab:sft_rl_comparison}
consolidates the results of CharDuplex-SFT and CharDuplex across character
consistency, general speech intelligence, and full-duplex interaction
capabilities. Because the metrics have different scales and optimization
directions, we report a direction-aware relative change:
\begin{equation}
\Delta_{\mathrm{rel}} =
\begin{cases}
\dfrac{x_{\mathrm{RL}}-x_{\mathrm{SFT}}}
      {x_{\mathrm{SFT}}}\times100\%, & \text{for } \uparrow \text{ metrics},\\[8pt]
\dfrac{x_{\mathrm{SFT}}-x_{\mathrm{RL}}}
      {x_{\mathrm{SFT}}}\times100\%, & \text{for } \downarrow \text{ metrics}.
\end{cases}
\label{eq:relative_change}
\end{equation}
Thus, a positive value consistently denotes improvement and a negative value
denotes degradation.

\begin{table*}[t]
\centering
\small
\setlength{\tabcolsep}{7pt}
\caption{
Consolidated comparison between CharDuplex-SFT and CharDuplex after FDGym-based
interactive post-training. Relative changes are direction-aware according to
Eq.~\ref{eq:relative_change}; positive values indicate improvement and negative
values indicate degradation. Percentages are computed from the reported scores
and may be affected slightly by rounding.
}
\label{tab:sft_rl_comparison}
\resizebox{\textwidth}{!}{
\begin{tabular}{llccc}
\toprule
Capability
& Metric
& CharDuplex-SFT
& CharDuplex
& Relative Change (\%) \\
\midrule

\multirow{5}{*}{Character consistency}
& IA $\uparrow$
& 0.807 & 0.815 & +0.99 \\
& CC $\uparrow$
& 0.800 & 0.805 & +0.63 \\
& PeC $\uparrow$
& 0.729 & 0.738 & +1.23 \\
& KC $\uparrow$
& 0.770 & 0.781 & +1.43 \\
& Avg. $\uparrow$
& 0.777 & 0.785 & +1.03 \\

\midrule

\multirow{10}{*}{General speech intelligence}
& LlamaQ $\uparrow$
& 74.3 & 74.7 & +0.54 \\
& WebQ $\uparrow$
& 40.8 & 40.5 & -0.74 \\
& OBQA $\uparrow$
& 63.1 & 62.4 & -1.11 \\
& MMSU $\uparrow$
& 44.2 & 44.0 & -0.45 \\
& SD-QA $\uparrow$
& 47.4 & 49.2 & +3.80 \\
& BBH $\uparrow$
& 50.0 & 51.2 & +2.40 \\
& AdvBench $\uparrow$
& 95.8 & 95.6 & -0.21 \\
& AlpacaEval $\uparrow$
& 3.63 & 3.63 & 0.00 \\
& CommonEval $\uparrow$
& 3.09 & 3.09 & 0.00 \\
& WildVoice $\uparrow$
& 2.30 & 2.29 & -0.43 \\

\midrule

\multirow{10}{*}{Full-duplex interaction}
& Pause Synthetic TOR $\downarrow$
& 0.934 & 0.927 & +0.75 \\
& Pause Candor TOR $\downarrow$
& 0.815 & 0.815 & 0.00 \\
& Backchannel TOR $\downarrow$
& 0.636 & 0.709 & -11.48 \\
& Backchannel ICC Freq. $\uparrow$
& 0.054 & 0.055 & +1.85 \\
& Backchannel JSD $\downarrow$
& 0.890 & 0.889 & +0.11 \\
& Smooth Turn Taking TOR $\uparrow$
& 0.840 & 0.857 & +2.02 \\
& Smooth Turn Taking Latency $\downarrow$
& 0.133 & 0.136 & -2.26 \\
& User Interruption TOR $\uparrow$
& 0.990 & 0.990 & 0.00 \\
& User Interruption GPT-4o $\uparrow$
& 4.081 & 4.071 & -0.25 \\
& User Interruption Latency $\downarrow$
& 0.558 & 0.565 & -1.25 \\

\bottomrule
\end{tabular}
}
\end{table*}

FDGym improves every SpeechRole-Eval metric, with consistent gains in
instruction adherence, conversational coherence, personality consistency, and
knowledge consistency. In contrast, the general speech-intelligence benchmarks
show small bidirectional fluctuations: improvements on some QA and reasoning
tasks are accompanied by similarly limited decreases on others, while the
open-ended response and safety metrics remain close to the SFT model.
Full-duplex interaction shows a similar overall pattern, with most metrics
remaining stable or changing only modestly, although Backchannel TOR exhibits
a larger relative decrease. Overall, the consolidated results indicate that
the primary effect of interactive post-training is stronger character
consistency, while the general speech intelligence and full-duplex interaction
capabilities established during supervised training are largely preserved.

\section{Ablation Studies and analysis}
\label{sec:app_ablation}
\paragraph{Effect of Timing Rewards}
We examine the contribution of the two deterministic timing rewards,
$r_{\mathrm{resp}}$ and $r_{\mathrm{yield}}$, used during FDGym-based RL.
We train an ablated variant by removing both timing terms while keeping the
LLM-judged rewards unchanged. Removing the timing rewards does not produce a meaningful improvement on
SpeechRole-Eval, while the full-duplex turn-taking performance
decreases slightly. This indicates that the timing rewards help preserve the full-duplex interaction behavior established during SFT. Since both
rewards are computed deterministically from the rollout timestamps and incur
negligible additional computational cost, we retain them in the final training
objective.

\paragraph{Effect of Dialogue Verification}
We evaluate whether the verifier in the automated data-construction pipeline
materially improves dialogue quality. We compare the complete
generator--verifier pipeline against an ablated variant that directly accepts
the generator output without verification or repair. For evaluation, we randomly
sample generated dialogues and use GPT-5.6 Sol\footnote{https://developers.openai.com/api/docs/models/gpt-5.6-sol} with maximum reasoning effort as
an independent judge. Without verification, the pass rate is 65.6\%. In the full generator--verifier pipeline, 68.8\%
of generated dialogues pass the initial verification directly, and the final pass rate, as judged by gpt-5.6-sol, increased to 93.7\%. Although removing verification
reduces data-construction cost, it substantially lowers the quality of the
resulting corpus and would require additional downstream filtering. We therefore
retain the generator--verifier process in the final data-construction pipeline.

\end{document}